\documentclass[aps,amsmath,nofootinbib,superscriptaddress,amssymb,preprintnumbers,floatfix,showpacs,preprint]{revtex4-1}
 \usepackage{amsmath,txfonts,longtable,booktabs,overpic,amssymb,bm,bbm,multirow,float,graphicx,color,dcolumn,subfigure,hyperref,tikz}
\definecolor{blue}{RGB}{45,48,146}
  \hypersetup{colorlinks,citecolor=blue,anchorcolor=red,menucolor=red, linkcolor=red,filecolor=red,runcolor=red,urlcolor=blue,frenchlinks=true}

\begin{document}

	\title{Effect of spatially oscillating field on Schwinger pair production}
	
	\author{Orkash Amat}
	\affiliation{ Key Laboratory of Beam Technology of the Ministry of Education, and College of Nuclear Science and Technology, Beijing Normal University, Beijing 100875, China}
	
	\author{Li-Na Hu}
	\affiliation{ Key Laboratory of Beam Technology of the Ministry of Education, and College of Nuclear Science and Technology, Beijing Normal University, Beijing 100875, China}

	\author{Mamat Ali Bake}
    \affiliation{ School of Physics Science and Technology, Xinjiang University, Urumqi 830046, China}

	\author{Melike Mohamedsedik}
	\affiliation{College of Xinjiang Uyghur Medicine, Hotan 848000, China}

	\author{B. S. Xie}\email{bsxie@bnu.edu.cn}
	\affiliation{ Key Laboratory of Beam Technology of the Ministry of Education, and College of Nuclear Science and Technology, Beijing Normal University, Beijing 100875, China}
	\affiliation{ Institute of Radiation Technology, Beijing Academy of Science and Technology, Beijing 100875, China
	}

\begin{abstract}
Effect of spatially oscillating fields on the electron-positron pair production is studied numerically and analytically when the work done by the electric field over its spatial extent is smaller than twice the electron mass. Under large spatial scale, we further explain the characteristics of the position and momentum distribution via tunneling time, tunneling distance and energy gap between the positive and negative energy bands in the Dirac vacuum. Our results show that the maximum reduced particle number is about five times by comparing to maximum number for non-oscillating field. Moreover, the pair production results via Dirac-Heisenberg-Wigner formalism can be also calculated by using local density approximation and analytical approximation method when spatial oscillating cycle number is large. Moreover, in case of large spatial scale field, the position distribution of created particles could be interpreted by the tunneling time.
\end{abstract}
\maketitle

\section{Introduction\label{sec:1}}

Schwinger effect is one of the fascinating nonperturbative phenomena in quantum electrodynamics (QED)\cite{Dirac:1928hu,Klein:1929zz,Sauter:1931zz,Heisenberg:1936nmg,Schwinger:1951nm}. This effect has not been yet observed directly in the laboratory as the critical field strength $E_{cr}=m^{2}c^{3}/e\hbar \thickapprox 1.3 \times 10^{16}\rm{V/cm}$ (corresponding laser intensity is about $4.3 \times 10^{29}\rm{W/cm^2}$, where $m$ and $-e$ are the electron mass and charge) is not feasible so far\cite{Alkofer:2001ik,Hebenstreit:2009km}. With the rapid development of the laser technology, however, the forthcoming laser intensity\cite{ELI,XCELS} is expected to reach $10^{24}-10^{26}\mathrm{W/cm^2}$ that has raised hopes of observing pair production in the future \cite{Abramowicz:2021zja}.

The effects of spacetime-dependent inhomogeneous fields on the pair production is an interesting issue for the studies in the strong field QED. We know that the different shapes of the spatial or temporal part of field have different effects on the pair production, e.g., temporal Sauter envelope\cite{Popov:2001ak}, spatial Sauter envelope\cite{Dunne:2005sx}, temporal super-Gaussian envelope\cite{Kohlfurst:2019mag}, spatial Gaussian envelope\cite{Ababekri:2019dkl} etc. Besides, the influence of different field parameters are also important, e.g., frequency chirp effect\cite{Mohamedsedik:2021pzb,Li:2021vjf,Hu:2022ouk} and phase effect\cite{Mohamedsedik:2022dxz}. For the non-plane wave background field, there are many studies for either some simple spatial inhomogeneous fields like of the cosine, Sauter and Gaussian shapes\cite{Dunne:2005sx,Kohlfurst:2019mag,Ababekri:2019dkl,Mohamedsedik:2021pzb,Li:2021vjf,Hu:2022ouk} or some time-dependent fields with complicated temporal shapes \cite{Li:2015cea,Kohlfurst:2017git,Olugh:2019nej,Ren:2023nqm}.

In Ref.\cite{Aleksandrov:2019ddt,Kohlfurst:2022edl}, the pair production in spatially oscillating fields is investigated, but spatial part is the cosine or sine function since effects of more complex spatial oscillation fields on pair production is still facing a theoretical challenging. For example, the momentum spectrum of the created pairs need more detailed examination for those complicated spatially oscillating inhomogeneous field \cite{Dumlu:2010ua,Dumlu:2011rr,Akkermans:2011yn}.  Among them there is a special case, in which the work done by the electric field over its spatial extent (we denote it $W$) may be smaller than twice of the electron mass. In this special case, however, the previous studies \cite{Schutzhold:2008pz,Hebenstreit:2011wk,Hebenstreit:2011,Kohlfurst:2015zxi} are focused mainly on the fields with the Gaussian-like shapes. To our knowledge, an effect of more complicated spatially oscillating inhomogeneous field on the pair production is still lacking enough research, therefore, it is necessary to study the effect of spatially oscillating field in the specific $W<2mc^2$.

Under such circumstance, we intend to study and answer some of involving problems. For example, what will happen to the distribution of pair production? Will the local density approximation (LDA) be applied to it? Is there any connection to existed rigorous solution analytically? and so on. On the other hand, the tunneling picture has played a key role to the vacuum pair production beside of the multiphoton mechanism for the temporal oscillating field when $\hbar \omega \leqslant 2mc^2$, where $\omega$ is the field frequency \cite{Schutzhold:2008pz}, it has been seen from the study for some interesting fields \cite{Amat:2022uxq}. Thus, it would be also helpful to understand the results of present research from the view point of tunneling time and distance.

Motivated by the factors mentioned above, therefore, in this paper, we investigate the vacuum pair production when $W<2mc^2$. First we give a simple perspective picture which reveals mainly the tunneling process. We further demonstrate its correctness for the pair production via different numerical approaches and analytical approximation. We interpret why and how the electron-positron pair can leave out from the vacuum. Moreover, we present and discuss the exact tunneling distance and time of the created particle, which is employed to understand the characteristics of the position and momentum distribution of created pairs. Finally, we show a relationship between the position distribution and tunneling time by employing the worldline instanton (WI) approach.

It is necessary to give a simple introduction to some of the widely used and powerful tools that can calculate particle distribution in spatial oscillating inhomogeneous fields. Among many methods to investigate pair production, as far as we know, some important ones include the WI technique \cite{Affleck:1981bma,Dunne:2006st,Dunne:2006ur,Dunne:2006ff,Dunne:2008zza,Dunne:2008kc,Dumlu:2011cc,BaisongXie:2012,Ilderton:2015lsa,Ilderton:2015qda,Schneider:2014mla,Xie:2017,Schneider:2018huk,Rajeev:2021zae,DegliEsposti:2021its,DegliEsposti:2022yqw}, the real time Dirac-Heisenberg-Wigner (DHW) formalism~\cite{Bialynicki-Birula:1991jwl,Olugh:2018seh,Li:2021wag,Kohlfurst:2021dfk,Kohlfurst:2021skr}, the computational quantum field theory ~\cite{Krekora:2004trv,PhysRevA.73.022114,Tang:2013qna,Wang:2019oyk,Wang:2021tmo}, the imaginary time method ~\cite{Popov:2005rp}, the quantum Vlasov equation (QVE)~\cite{Kluger:1998bm,Schmidt:1998vi,Li:2014xga,Li:2014psw,Gong:2020jqs}, the Wentzel-Kramers-Brillouin (WKB) approach~\cite{Popov:1971iga,Kim:2007pm,Strobel:2014tha,Oertel:2016vsg}, scattering matrix approach ~\cite{Ritus:1985,Titov:2018bgy,Tang:2019ffe,Ilderton:2019ceq,King:2019igt,Seipt:2020diz,Wistisen:2020rsq,Lei:2021eqe,Podszus:2021lms,Bu:2021ebc,Tang:2022tmn,Golub:2022cvd,MacLeod:2022qid}, and so on. In particular, the DHW formalism allows us to investigate the pair production for any background field, not limited to plane wave~\cite{Kohlfurst:2019mag}. On the other hand, due to the complex nature of the pair production under spacetime-dependent field, one can find only a few analytical results for simple background modes~\cite{Narozhnyi:1970,Brezin:1970xf,Hebenstreit:2010vz,Dunne:2012,Amat:2022uxq}. Hence, we have to adopt numerical methods to study various natures of pair production. In present work, we will use the DHW formalism as numerical approach.

The paper is organized as follows. In Sec.~\ref{sec:2}, we briefly introduce the vacuum description of pair production process in spatial oscillating field. In Sec.~\ref{sec:3}, numerical and analytical approaches are introduced. In Sec.~\ref{sec:4}, we derive the analytical solution of the tunneling time for large spatial scale via the WIF. In Sec.~\ref{sec:5}, we give and discuss our numerical and analytical results, and interpret the momentum and position distributions of created pairs. Meanwhile, we show the relation between the tunneling time and position distribution, and explain the position distribution. Finally, the summary is given briefly in Sec.~\ref{sec:6}.

We use the natural units ($\hbar=c=1$), throughout this paper, and express all quantities in terms of the electron mass $m$.

\section{Pair production process in spatial oscillating field}\label{sec:2}

\begin{figure}[ht!]\centering
\includegraphics[width=0.45\textwidth]{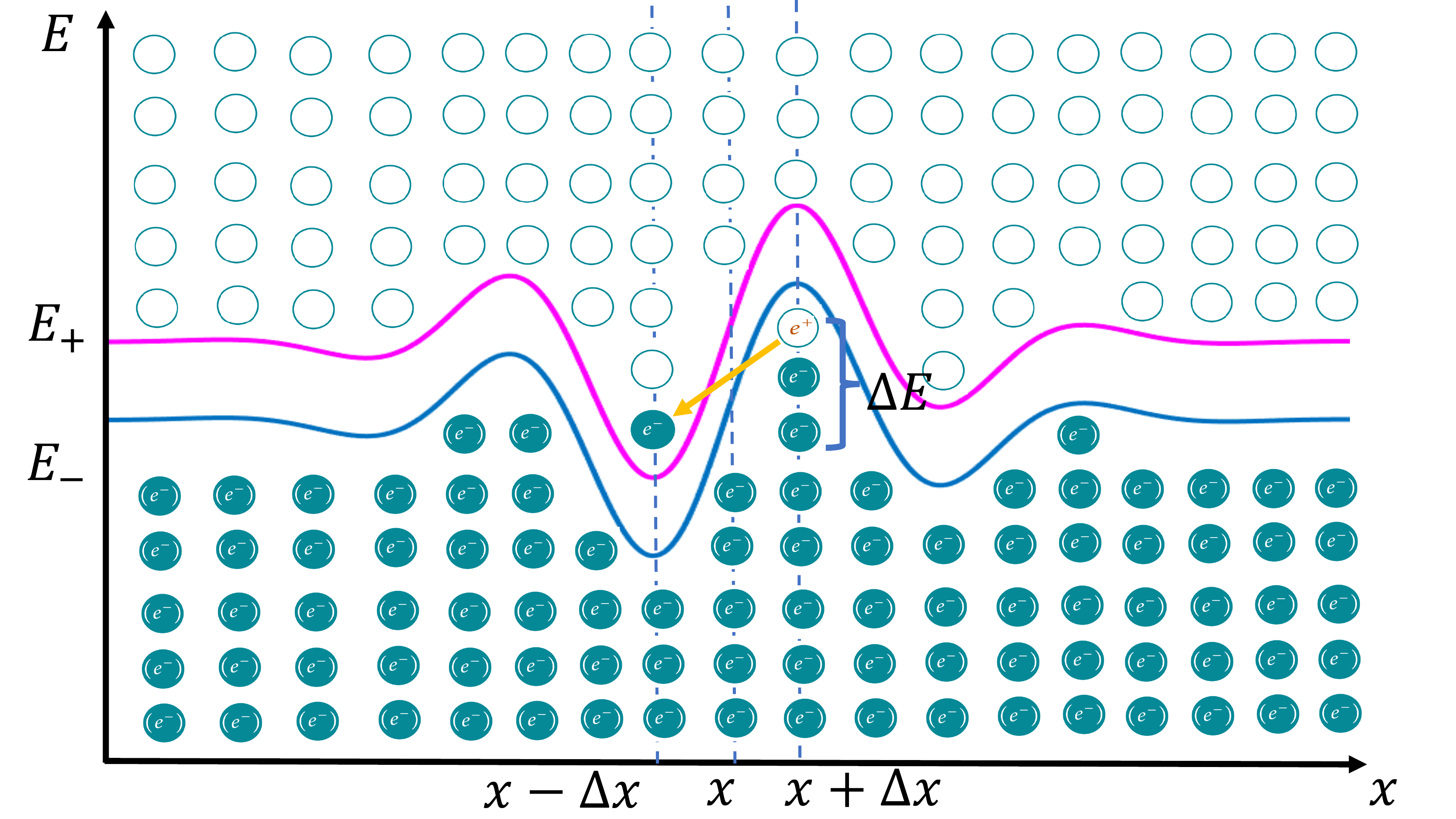}
\caption{Plot for the Dirac vacuum under high spatial oscillating spacetime-dependent electric field, where $E_{\pm}(x)=\pm m c^2 + \phi(x)$ when $W < 2 m c^2$. The magenta and blue lines denote the $E_{+}(x)$ (minimum positive energy band) and $E_{-}(x)$ (maximum negative energy band) for the same center-of-mass coordinate $x$ respectively.
}\label{fig:1}
\end{figure}

The electron-positron will leave out of the vacuum after the particle jumps from any negative high energy state at $x_{+}=x+\Delta x$ to any positive low energy state at $x_{-}=x-\Delta x$ in the Dirac semiclassical picture of the vacuum when the work $W=e\int \mathbf{E}\cdot d\mathbf{x}<2mc^2$ as shown in Fig.~\ref{fig:1}. The electron position distribution depends on the center-of-mass coordinate $x$~\cite{Bialynicki-Birula:1991jwl}, thus, we will consider above transition probability of the particle at $x$ in order to interpret the position and momentum distributions of electron. Moreover, we can obtain the transition (tunneling) distance $d=x_{+}-x_{-}=2 \Delta x$, see Fig.~\ref{fig:1}. Note that this quantum jumping process includes the tunneling process under an external spatial oscillating spacetime-dependent field. The energy gap between the two energy bands $\Delta E$ is transformed into the created pair's energy. From energy conservation of this process, $\Delta E = E_{e^-} + E_{e^+}$, we can find the general relativistic relation under the pure external field (does not include the pondermotive force~\cite{Hebenstreit:2011pm,Hebenstreit:2011wk,Kohlfurst:2015zxi}) as
\begin{align}\label{eq:1}
\left(\frac{\Delta E}{2}\right)^2=m_{*}^{2}+q^{2},
\end{align}
where $m_{*}$ is effective mass~\cite{Kohlfurst:2013ura}, $q$ denotes a characteristic residual kinetic momentum of the outgoing particles from the tunneling process. According to our new perspective picture, the electron-positron pair can escape successfully from the vacuum by jumping from the negative high energy state to the positive low energy state when $W<2mc^2$. This is why the electron-positron pairs can get out of the vacuum.

In order to better understand, we consider an example of  spacetime-dependent spatially oscillating electric field. In our case, we ignore particle momenta orthogonal to this dominant direction. We choose $A^\mu\left(x,t\right) =\left( \phi\left(x\right) f(t),0,0,0 \right)$. The electric field can be written in the following form
\begin{align}\label{eq:2}
&E\left(x,t\right)=\varepsilon E_{cr}g(x)f(t),
\end{align}
where $g(x)={\rm cos}\left(k x\right)e^{-\frac{x^2}{2\lambda^2}}$ and $f(t)={\rm sech}^2\left(\omega t\right)$, $\varepsilon$ is the peak field strength, $E_{cr}$ denote the critical field strength for electric and magnetic field, $\lambda$ is spatial scale, $k$ is the spatial wave number, $\sigma_{\lambda}=k\lambda$ is spatial oscillating cycle number. Accordingly, the corresponding scalar potential is
\begin{align}\label{eq:3}
&\phi(x)= -\frac{\sqrt{\pi}\lambda}{\sqrt{8}}e^{-\frac{1}{2}k^2\lambda^2}\left({\rm erf}\left(\frac{x-ik \lambda^2}{\sqrt{2}\lambda}\right)+ {\rm erf} \left(\frac{x+ik \lambda^2}{\sqrt{2}\lambda} \right)\right).
\end{align}

\begin{figure}[ht!]\centering
\includegraphics[width=0.45\textwidth]{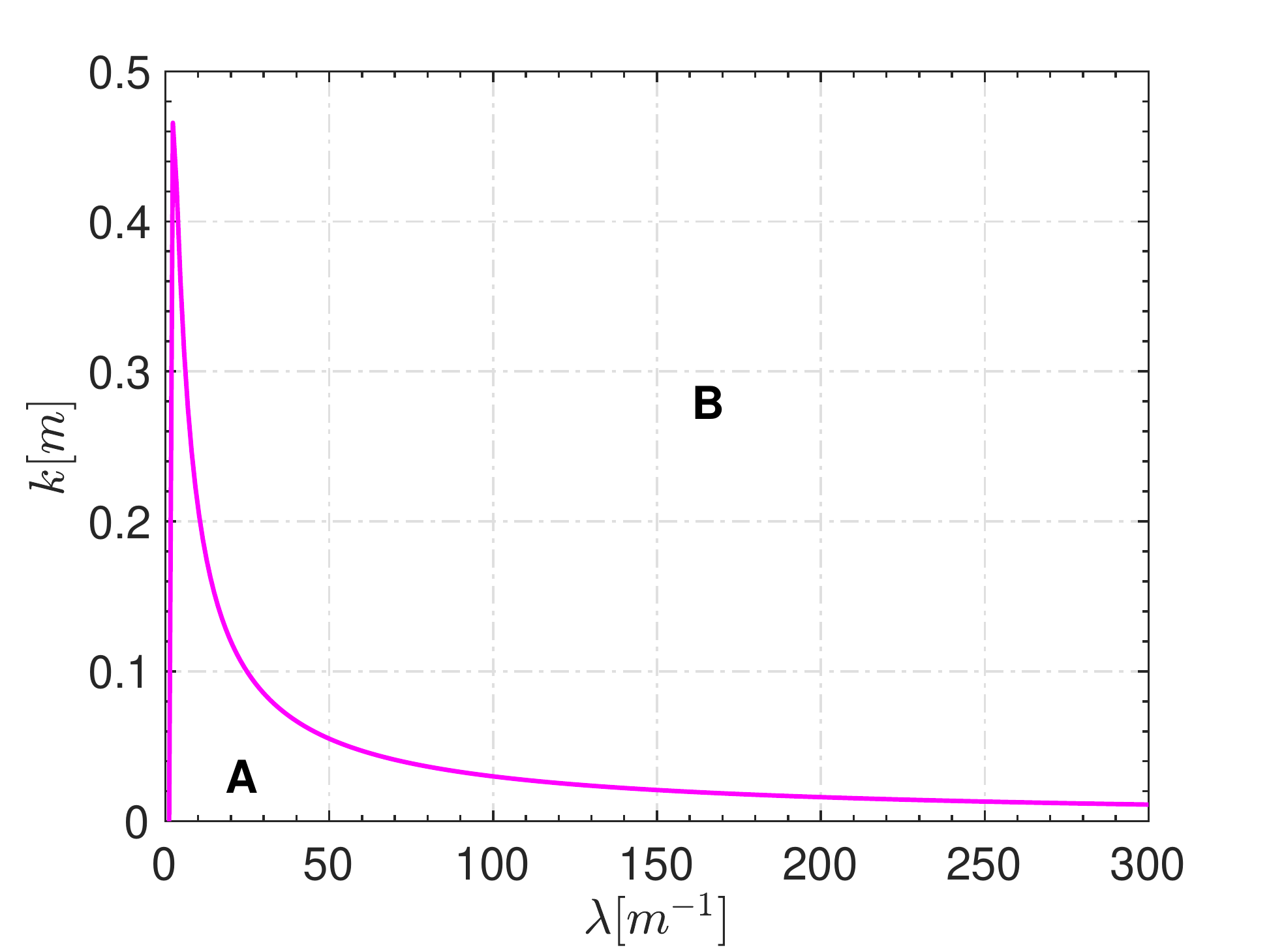}
\caption{Plot of the work $W$ for $k$ and $\lambda$. Areas A and B represent $W \geqslant 2mc^2$ and $W<2mc^2$ respectively. The magenta line denotes for $W=2mc^2$.
\label{fig:2}}
\end{figure}

To better compare optimization schemes, we consider the external field which is the same energy in the spacetime with different $k$, $\lambda$ and $\omega$. We define $\varepsilon_{0}=\varepsilon=0.5$ for $k=0$ case. The total energy of the external field in the same 2-space volume $V_2$ can be written as
\begin{align}\label{eq:4}
\mathcal{E} =\frac{V_2}{2} \int\int E^{2}(x,t)dxdt={\rm constant},
\end{align}
If we fix the $\omega$ in the whole work, we can obtain the peak field strength for different $k$ as
\begin{align}\label{eq:5}
\varepsilon = \sqrt{\frac{2}{1+e^{-k^2\lambda^2}}}\varepsilon_{0}.
\end{align}
In Fig.~\ref{fig:2}, areas A and B represent $W \geqslant 2mc^2$ and $W<2mc^2$ cases respectively. Note that the vacuum pair production in the area A is well understood in many previous works~\cite{Hebenstreit:2011wk,Hebenstreit:2011,Kohlfurst:2015zxi}, but the pair creation in area B need to further study. Hence, in the present paper we would like to focus specifically in the area B to investigate the vacuum pair production when $W<2mc^2$.

\section{Numerical and analytical methods}\label{sec:3}

\subsection{DHW formalism}\label{sec:A}

For studying the electron-positron pair production of vacuum in the background fields we write the Lagrangian~\cite{Hebenstreit:2011}
\begin{align}
\label{eq:6}
\mathcal L \left(\Psi,\bar{\Psi},A \right)=\frac{i}{2} \left(\bar{\Psi} \gamma^{\mu}\mathcal{D}_{\mu}\Psi-\bar{\Psi}\mathcal{D}_{\mu}^{\dag} \gamma^{\mu}\Psi\right)-m\bar{\Psi}\Psi-\frac{1}{4}F_{\mu \nu} F^{\mu \nu},
\end{align}
where $\mathcal{D}_{\mu}=\left(\partial_{\mu}+ieA_{\mu}\right)$ is the covariant derivative and correspondingly $\mathcal{D}_{\mu}^{\dagger} = (\overset{\leftharpoonup}{\partial_{\mu}}-ieA_{\mu})$. In order to describe the dynamics of the particles, we need the Dirac equation
\begin{align}
\label{eq:7}
\left(i\gamma^{\mu}\partial_{\mu}-e\gamma^{\mu}A_{\mu}(r)-m\right)\Psi (r)=0,
\end{align}
and the adjoint Dirac equation
\begin{align}
\label{eq:8}
\bar{\Psi}(r)\left(i\gamma^{\mu}\overset{\leftharpoonup}{\partial_{\mu}} +e\gamma^{\mu}A_{\mu}(r)+m\right)=0.
\end{align}
The Dirac spinors $\Psi$, $\bar{\Psi}$ and the vector potential $A^{\mu}\left(r\right)$ are the main ingredients in the DHW formalism. Note that, the background is considered to be a classical one. Further, we introduce the density operator as ~\cite{Hebenstreit:2011}
\begin{align}\label{eq:9}
 \hat{\mathcal C}_{\alpha \beta}\left(r,s\right) =\mathcal U \left(A,r,s
\right) \ \left[\bar \Psi_\beta \left(r-s/2\right),\Psi_\alpha \left(r+s/2\right)\right],
\end{align}
where $r$, $s$ denotes the center-of-mass and the relative coordinate of two particles.
The Wilson line factor is used to make the density operator gauge invariant under the $U(1)$ gauge:
\begin{equation}
\label{eq:10}
 \mathcal U \left(A,r,s\right)=\exp\left(\mathrm{i}es\int_{-1/2}^{1/2}d\xi A\left(r+\xi s\right)\right).
\end{equation}

In order to perform numerical calculations, we use the DHW as the powerful tool in our study due to it allows us to investigate vacuum pair production for  inhomogeneous field ~\cite{Bialynicki-Birula:1991jwl}. This method is well applied to the case of one spatial dimension, in which there are only four Wigner components, $\mathbb{S}$, $\mathbb{V}_{0}$, $\mathbb{V}_x$ and $\mathbb{P}$ for electric field $E(x,t)$.
The DHW equations of motion in this case of 1+1 can be written as~\cite{Hebenstreit:2011wk}
\begin{align}
\label{eq:11}
	&D_t \mathbbm{S}-2p_x \mathbbm{P}=0, \\ \label{eq:12}
	&D_t \mathbbm{V}_0 +\partial_x\mathbbm{V}_x=0,\\ \label{eq:13}
	&D_t \mathbbm{V}_x +\partial_x\mathbbm{V}_0=-2\mathbbm{P},\\ \label{eq:14}
	&D_t \mathbbm{P}+2p_x \mathbbm{S}=2m \mathbbm{V}_x,
\end{align}
where
\begin{equation}
  \label{eq:15}
	D_t=\partial_t + e\int_{-1/2}^{1/2}{d\xi E_x(x+i\xi\partial_{p_x},t)\partial_{p_x}}.
  \end{equation}

In order to perform simulation more conveniently, we can define 4 Wigner components as $\mathbb{W}_{0}=\mathbb{S}$, $\mathbb{W}_1=\mathbb{V}_0$, $\mathbb{W}_2=\mathbb{V}_x=\mathbb{V}$ and $\mathbb{W}_3=\mathbb{P}$. So from the initial conditions given by the vacuum solution for single particle
\begin{equation}
\label{eq:16}
\mathbb{S}_{vac} \left(p_x \right)=-\frac{2}{\sqrt{1+p_x^2}},\
\mathbb{V}_{vac} \left(p_x \right)= -\frac{2p_x}{\sqrt{1+p_x^2}},
\end{equation}
we have the modified Wigner components as
  \begin{equation}
   \label{eq:17}
   \mathbb{W}_i^v=\mathbb{W}_i - \mathbbm{W}_{i,vac}.
  \end{equation}
Finally, we can calculate the particle number density at asymptotic times $t_f \to +\infty$
  \begin{align}
   \label{eq:18}
  n \left(x,p_x,t\to +\infty \right)=\frac{\mathbb{S}^v+p_x \mathbbm{V}^{v}_x}{\sqrt{1+p_x^2}}.
  \end{align}
The momentum and position distributions are given by
  \begin{align}
   \label{eq:19}
   n\left(p_x,t\to +\infty\right)&=\int \frac{dx}{2\pi}n\left(x,p_x,t\to+\infty\right), \\ \label{eq:20}
   n \left(x,t\to +\infty\right)&=\int dp_x n\left(x,p_x,t\to +\infty \right).
  \end{align}
And the total particle number is readily got as
  \begin{align}
   \label{eq:21}
   N(t\to +\infty)&=\int{\rm d}p_x n\left(p_x,t\to +\infty\right).
  \end{align}
It is worthy to point out that for the convenient comparison we should cope with the reduced quantities $\bar{n}(p_x,t \to +\infty)=n(p_x,t \to +\infty)/\lambda$ and $\bar{N}(t \to +\infty)= N(t \to +\infty)/\lambda$ under the same energy.

\subsection{LDA}\label{sec:C}

When the spatial variation scale is much larger than the Compton wavelength $\lambda\gg \lambda_C$, the vacuum pair production can be locally described by fixing point $x$~\cite{Hebenstreit:2011wk}. If we assume $E\left(x,~t\right)=\varepsilon E_{cr} g\left(x\right) h\left(t\right)$ and can $\varepsilon\left(x\right)=\varepsilon E_{cr} g\left(x\right)$ as effective field strength, where $ h\left(t\right)$ is an arbitrary time-dependent function. We can obtain momentum and positron distributions by summing results for homogeneous fields with different field strengths given as~\cite{Hebenstreit:2011wk}
\begin{align}
\label{eq:22}
\bar{n}_{loc} \left(p_x,t\to +\infty \right)&=\int\frac{dx}{2\pi}\bar{n}_{loc}\left(\varepsilon\left(x\right)|p_x, t\to +\infty\right), \\ \label{eq:23}
\bar{n}_{loc}\left(x, t\to +\infty\right)&=\int dp_x \bar{n}_{loc} \left( \varepsilon\left(x\right)|p_x, t\to +\infty\right).
\end{align}
$\bar{n}_{loc} \left( \varepsilon\left(x\right)|p_x, t\to \infty\right)$ can be found by using quantum kinetic theory at any fixed point $x$ for a time-depend electric field $E\left(t\right)=\varepsilon\left(x\right) h\left(t\right)$~\cite{Hebenstreit:2011wk}.

We choose time-dependent spatial homogeneous DHW method to find one particle distribution~\cite{Blinne:2015zpa}, and calculate the LDA result. The one-particle momentum distribution function $n(\mathbf{p},t)$ can be obtained by solving the following ten ordinary differential equations and the nine auxiliary quantities $\mathsf{V}_i(\mathbf{p},t):=\mathbbm{V}_i(\mathbf{p},t)$, $\mathsf{A}_i(\mathbf{p},t):=\mathbbm{A}_i(\mathbf{p},t)$ and $\mathsf{T}_i(\mathbf{p},t):=\mathbbm{T}_i(\mathbf{p},t)$~\cite{Blinne:2015zpa}:
\begin{align}\label{eq:24}
\begin{split}	
\dot{n}&=\frac{e}{2\Omega} \, \,  \mathbf{E}\cdot \mathsf{V},\\
\dot{\mathsf{V}}&=\frac{2}{\Omega^{3}} \left( \left(e\mathbf{E}\cdot \mathbf{p}\right)\mathbf{p}-e\Omega^{2}\mathbf{E}\right) (n-1)-\frac{\left(e\mathbf{E}\cdot \mathsf{V}\right)\mathbf{p}}{\Omega^{2}}-2\mathbf{p}\times \mathsf{A} -2m \mathsf{T},\\
\dot{\mathsf{A}}&=-2\mathbf{p}\times \mathsf{V},\\
\dot{\mathsf{T}}&=\frac{2}{m}\left[m^{2}\mathsf{V}+\left(\mathbf{p}\cdot \mathsf{V}\right)\mathbf{p}\right].
\end{split}
\end{align}
Initial condition values are selected as $n(\mathbf{p},-\infty)=\mathsf{V}(\mathbf{p},-\infty)= \mathsf{A}(\mathbf{p},-\infty)=\mathsf{T}(\mathbf{p},-\infty)=0$ in order to perform the calculation.  We can further obtain the one-particle momentum distribution $n(\mathbf{p},t)$.

It is worthy to note that, in our simulation, the Runge-Kutta method of 8(5,~3) order is used in order to avoid unphysical results during numerical calculation, in which we used ${\rm  RelTol= AbsTol}=10^{-10}$ (where we have specified a relative RelTol as well as an absolute error tolerance AbsTol). In order to calculate the various distribution with high accuracy, the lattice sizes have been set to $N_x \times N_{p_x}=8192\times4096$ and $N_x \times N_{p_x}=16384\times4096$ for low and high spatial oscillating field. Note that $\mathbf{p} \left(t=-\infty \right)=\mathbf{p} \left(t=+\infty \right)= \mathbf{q}$ in the time-depended Sauter pulse.

\subsection{Analytical approximation for large spatial scale} \label{sec:B}

The result can be obtained analytically by replacing the field strength $\varepsilon$ in the analytical one-particle distribution solution with an effective field strength $\varepsilon\left(x\right)=\varepsilon E_{cr}g(x)$ in spacetime-depended field when the field spatial scale is large. For example, we can get the analytical solution explicitly for $E\left(x,t\right)=\varepsilon E_{cr}g\left(x\right){\rm sech}^{2}\left(\omega t\right)$ by replacing $\varepsilon$ in the QVE solution for $E\left(t\right)=\varepsilon~{\rm sech}^{2}\left(\omega t\right)$ ~\cite{Hebenstreit:2011pm} with $\varepsilon\left(x\right)$ as (refer to Eq.~(99) of Ref.~\cite{Hebenstreit:2010vz})
\begin{widetext}
\begin{equation}
\label{eq:25}
n\left(x,p_x,t\rightarrow +\infty\right)=\frac{2\sinh \left(\frac{\pi}{2\omega}[\frac{2e}{\omega}\varepsilon\left(x\right)+\Tilde{Q}\left(x\right)-Q\left(x\right)]\right)\sinh\left(\frac{\pi}{2\omega}[\frac{2 e}{\omega}\varepsilon\left(x\right)-\Tilde{Q}\left(x\right)+Q\left(x\right)]\right)}
{
\sinh\left(\frac{\pi}{\omega}\Tilde{Q}\left(x\right)\right)
\sinh\left(\frac{\pi}{\omega}Q\left(x\right)\right)},
\end{equation}
\end{widetext}
where
\begin{align}\label{eq:26}
\Tilde{Q}\left(x\right)&=\sqrt{m^2 + p_{\bot}^2 + \left(p_x+ \frac{2e}{\omega} \varepsilon\left(x\right)\right)^2},
\end{align}

\begin{align} \label{eq:27}
Q\left(x\right)&=\sqrt{m^2+p_{\bot}^2+\left(p_x-\frac{2e}{\omega} \varepsilon\left(x\right)\right)^2}.
\end{align}
In the following we denote this treatment as the analytical approximation (AA) for large spatial scale approach.

\section{Tunneling time for large spatial scale}\label{sec:4}

To interpret explicitly the features of the position distribution, one needs to introduce tunneling time for spacetime-depended inhomogeneous field when the spatial variation scale is much larger than the Compton wavelength $\lambda\gg \lambda_C$ (slowly-varying-envelope approximation). The relationship between tunneling time in the Minkowski space and Euclidean space has been investigated in Ref.~\cite{Amat:2022uxq}. $T_{t}$ is the tunneling time for any spacetime-depended inhomogeneous fields, $x_{\pm}$ are the classical turning points, $x_{4}^{min}$ and $x_{4}^{max}$ are maximum and minimum of the fourth WI path $x_{4}$ in the Euclidean space, $\phi(x)$ is the potential of the field. From Ref.~\cite{Amat:2022uxq}, we can know the definition of the tunneling time, the time taken by the particle from $x_{-}$ to $x_{+}$ in the barrier region is tunneling time (quantum tunneling time). Further, we can achieve the tunneling time easily by performing Wick-rotation in order to simplify the path integral via $x_{4}=it$, see Refs.~\cite{Dunne:2005sx}. The tunneling time can be written as ~\cite{Amat:2022uxq}
\begin{align}
\label{eq:28}
&T_{t}=2\left(x_{4}^{max}-x_{4}^{min}\right).
\end{align}
If we choose the single-pulse time-dependent electric background
\begin{align}
\label{eq:29}
E(t)=E{\rm sech}^2(\omega t).
\end{align}
The WI paths can be obtained in the scalar or spinor QED~\cite{Dunne:2005sx}
\begin{align}
\label{eq:30}
x_3(u)&=\frac{m}{eE}\frac{1}{\gamma \sqrt{1+\gamma^2}}{\rm arcsinh}\left[\gamma\cos\left(2\pi u\right)\right], \\ \label{eq:31}
x_4(u)&=\frac{m}{eE}\frac{1}{\gamma}\arcsin\left[\frac{\gamma}{\sqrt{1+\gamma^2}}\sin\left(2\pi u\right)\right],
\end{align}
where $u=\tau/T$, in which $\tau$ and $T$ denote the proper-time and period of the WI path, $\gamma=m\omega/eE$. Due to lack of analytical solution, it is hard to obtain the analytical solution of the tunneling time directly under spacetime-depended inhomogeneous field. When the spatial variation scale is much larger than the Compton wavelength $\lambda\gg \lambda_C$, we can use spatial slowly-varying-envelope approximation. Thus, the tunneling time for spacetime-dependent field can be locally described by replacing the $\varepsilon$ in the analytical tunneling time solution under time-dependent field with $\varepsilon(x)=\varepsilon E_{cr}g(x)$. The tunneling time for our field could be obtained as
\begin{align}\label{eq:32}
T_{t}\left(x\right)=\frac{2}{\omega}{\rm arcsin}\left[\frac{\gamma(x)}{\sqrt{1+\gamma^{2}(x)}}\right],
\end{align}
where
\begin{align}\label{eq:33}
\gamma(x)=\frac{m~\omega}{e|\varepsilon\left(x\right)|}.
\end{align}
Note that the tunneling time was obtained by using the WI technique~\cite{Dunne:2005sx}, and obviously it is based on the Bohm viewpoint~\cite{Landauer:1994}.

\section{Results}\label{sec:5}

Now, we begin to prove the correctness of our perspective picture by adopting numerical and analytical methods. Here we use the DHW formalism, LDA and AA approaches.

\begin{figure}[ht!]\centering
\includegraphics[width=0.45\textwidth]{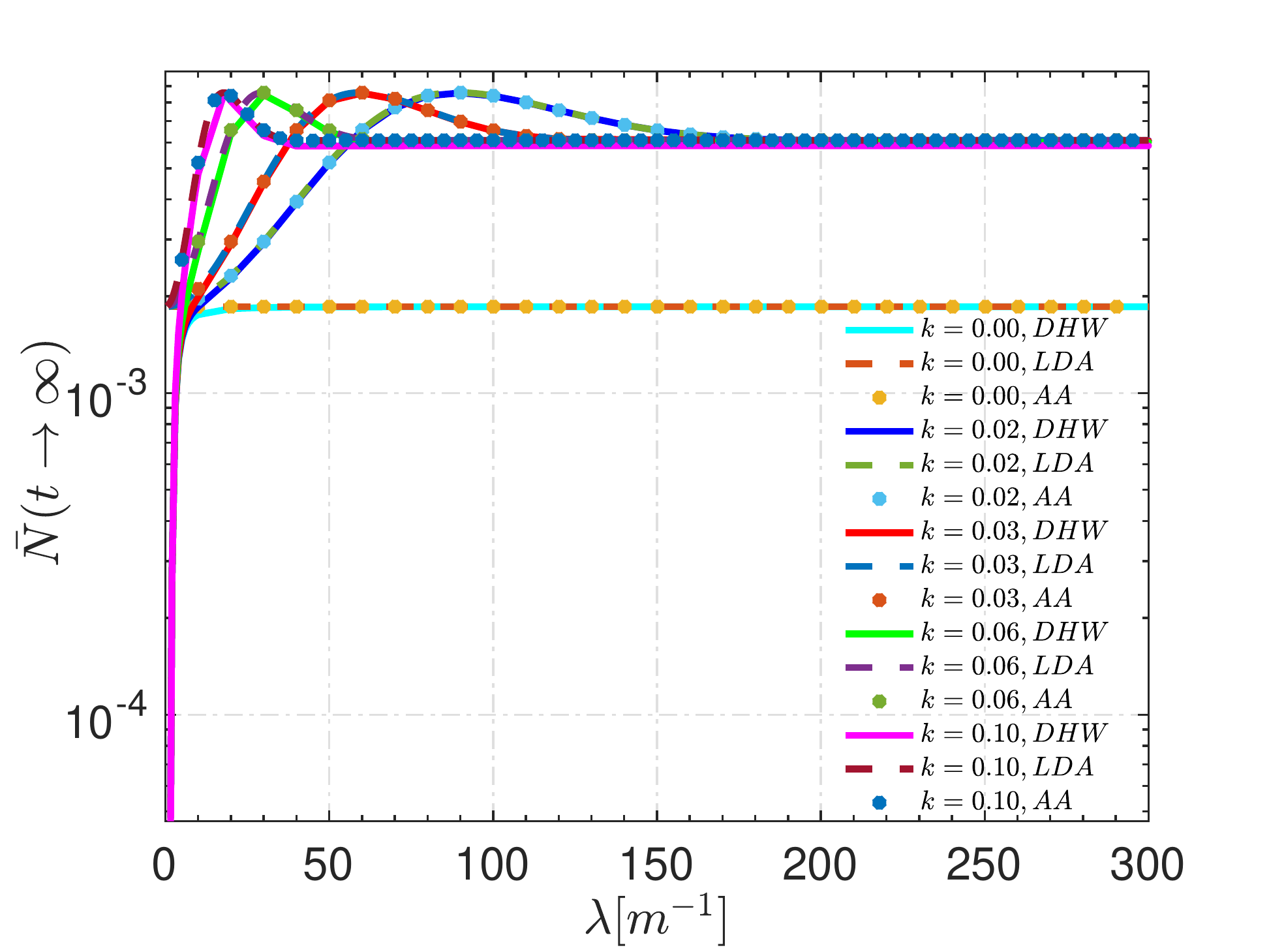}
\caption{Plot of the reduced total particle number for the DHW (solid lines), LDA (dashed lines) and AA (dotes) approaches with various $k$ and $\lambda$ when $\omega=0.1~m$.
\label{fig:3}}
\end{figure}

The reduced total particle number achieved by the DHW, LDA and AA approaches is shown as Fig.~\ref{fig:3}. Our results show that the maximum reduced particle number is about five times by comparing to that of \cite{Hebenstreit:2011wk}, meanwhile, the maximum number corresponds to the parameter regime that belongs to area A. Interestingly, however, the particle number when $W<2mc^2$ is still larger than normal case ($k=0$) in Fig.~\ref{fig:3}. On the other hand, the particle number when $k\neq 0$ is always larger than that when $k=0$, by the way, in which and area A, we would recover the result to the Fig.2 of Ref.~\cite{Hebenstreit:2011wk}. Furthermore, the total  particle distributions and maxima obtained from the three different approaches are approximately the same for appropriate $\lambda$. For example, when $k=0.1m$, the results of the three approaches are approximately the same for $5m^{-1}\leqslant \lambda \leqslant +\infty$. This means that the spatial oscillating effect on the pair production can be tender when spatial oscillating cycle number is larger than two so that the LDA and AA are valid.

\begin{figure}[ht!]\centering
\includegraphics[width=0.45\textwidth]{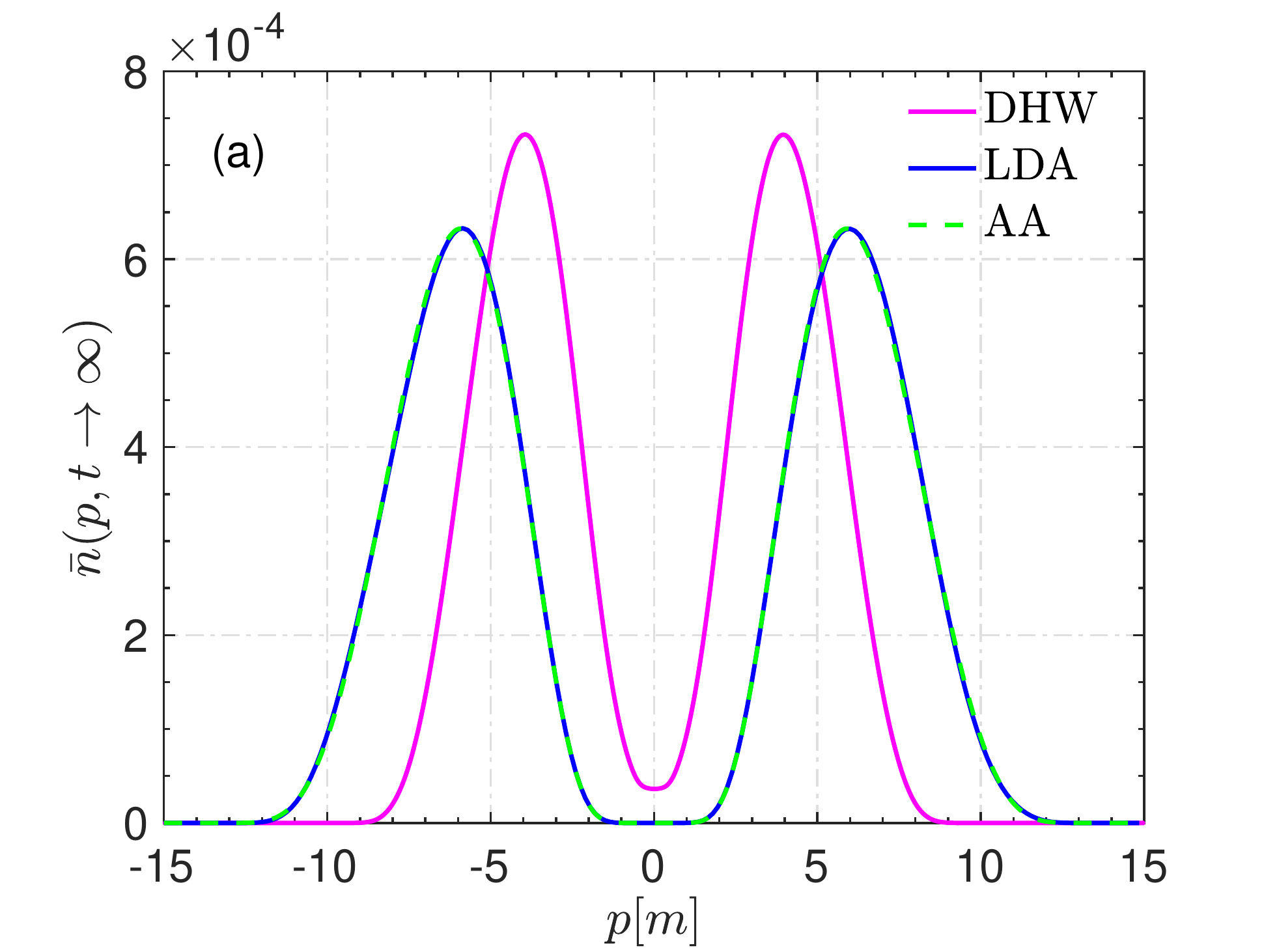}
\includegraphics[width=0.45\textwidth]{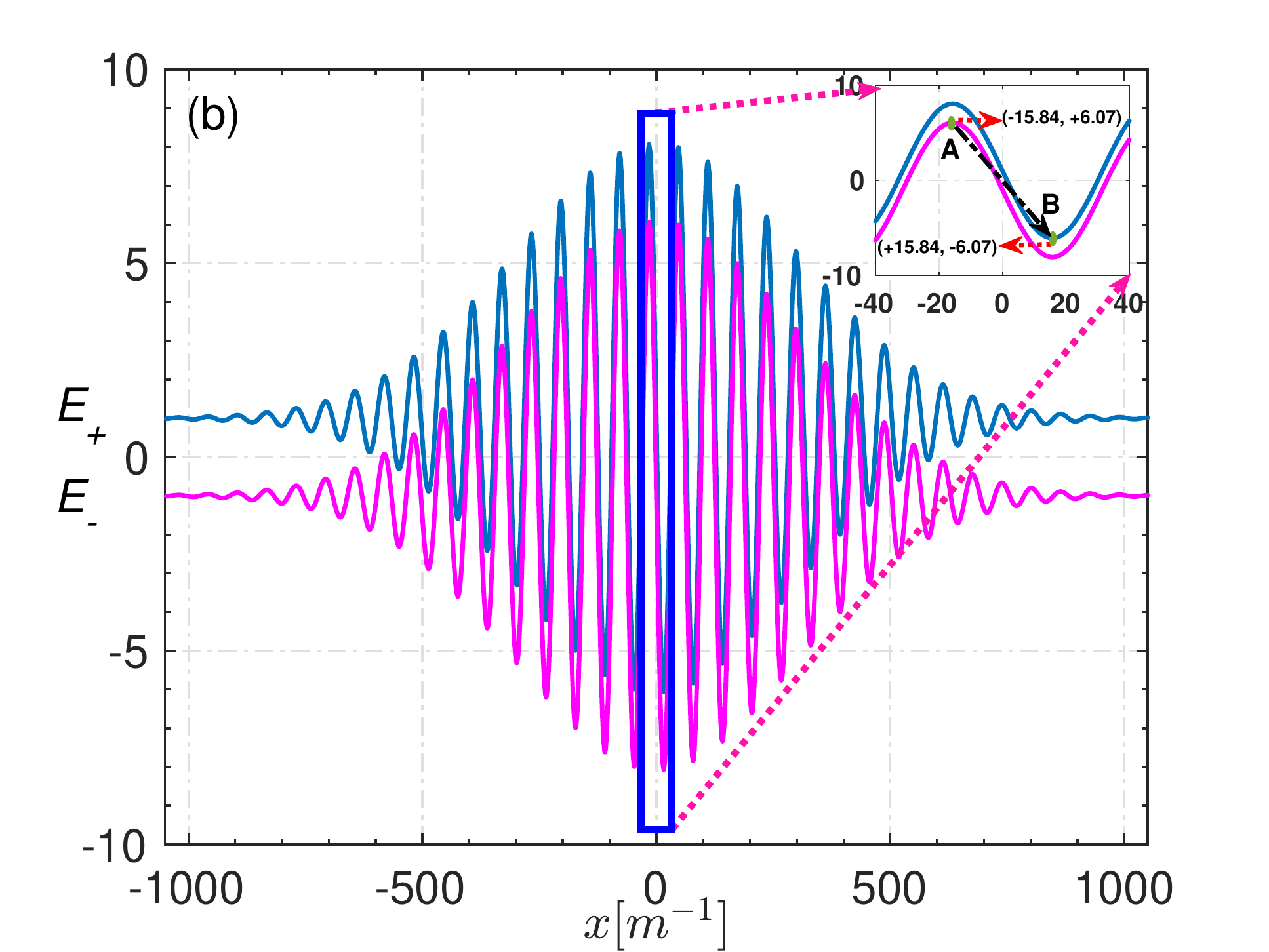}
\includegraphics[width=0.45\textwidth]{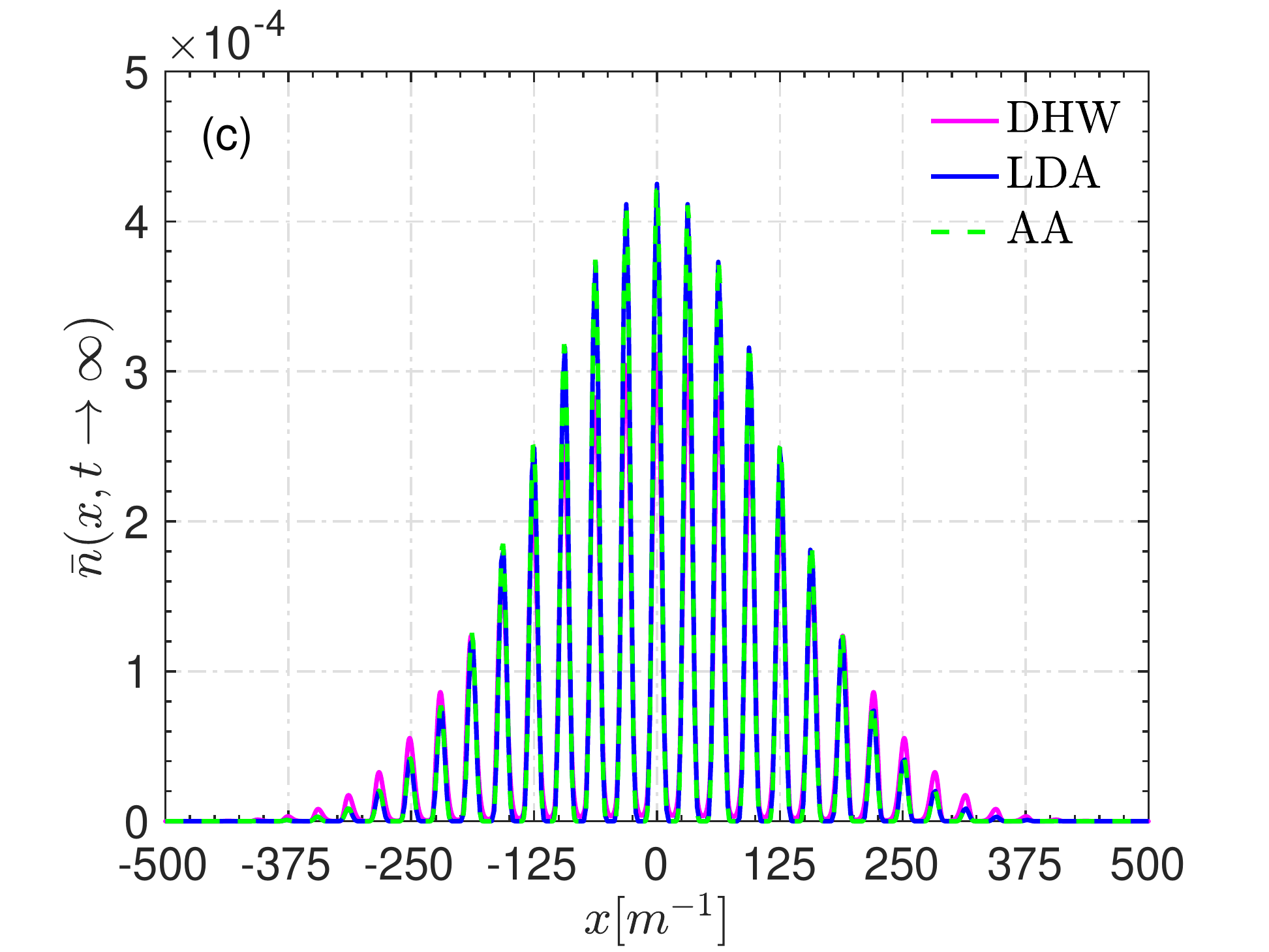}
\caption{Plots of momentum distribution (a), Dirac vacuum (b) and position distribution (c) when $W<2mc^2$. The magenta, blue and green dashed lines represent the DHW, LDA and AA approaches, respectively, when $k=0.1~m$, $\lambda=300~m^{-1}$ and $\omega=0.1~m$.
\label{fig:4}}
\end{figure}

Although the approximate same reduced total particle number is obtained by the different three methods of DHW, LDA and AA, it does not mean that the created pair experiences the same physical process. To see their differences, the momentum distribution is plotted in Fig.~\ref{fig:4}(a). While different approaches have different momentum distributions, they have almost the same area, which leads to the same total particle numbers approximately. Now we can interpret it by using our perspective picture mentioned in Sec.~\ref{sec:2}, for example, the momentum corresponding to the maximum of the momentum distribution. As shown in Fig.~\ref{fig:4}(b), one notes that the electron in the negative energy state jumps from the point A to the point B, and during this process, the energy gap $\Delta E$ between A and B points would transform the energy to the created pair. At the same time, the transition probability of the electron is the largest because the transition probability is proportional to the energy gap $\Delta E$ and inversely proportional to the transition, i.e. tunneling, distance $d=x_{B}-x_{A}=2\Delta x=\pi/k$, where $x_{B}$ and $x_{A}$ are the positions of $A$ and $B$ points in Fig.~\ref{fig:4}(b). Thus, the maximal energy gap could be found as $\Delta E=E_{B}-E_{A}\approx 12.1014~m$, where $E_{B}$ and $E_{A}$ denotes the energy for $A$ and $B$ points in Fig.~\ref{fig:4}(b), respectively. Then we can obtain $p\approx \pm 5.96749~m$ appropriately for $x=0$ point by using Eq.~\eqref{eq:1}. Surprisingly this value is just appropriately the momentum corresponding to the maximum of the momentum distribution for LDA formalism, i.e., $p_{peak}\approx \pm 5.97215~m$ in Fig.~\ref{fig:4}(a). We stress that although the momentum distributions of the LDA and AA methods are exactly the same shape, but the LDA and AA methods do not include the charge density as comparable to the DHW formalism where the charge density is present.

Another interesting feature is for the particle position distribution, shown in Fig.~\ref{fig:4}(c), could be also understood via our perspective picture. From it one can find that the strong oscillation occurs in the position distributions. To see the oscillatory phenomenon more clearly, we come back to the Fig.~\ref{fig:4}(b) again and have an intuitive looking at the maxima and minima transition probability at the center-of-mass coordinate $x$ via our perspective picture. The essentials of these oscillations is the connection between the distance for transition probability from the maxima to minima around of $x=M\pi/k$, where $M=0,\pm 1,\pm 2,\pm 3, ...$, which corresponds to jumping from peak to trough in Fig.~\ref{fig:4}(b), and the tunneling distance $d=2\Delta x=\pi/k$. Obviously these two distances are the same. Similarly, around of $x=(1/2+M)\pi/k$, which corresponds to jumping from trough to peak in Fig.~\ref{fig:4}(b), the jump-transition has the same tunneling distance $d=2\Delta x=\pi/k$. This is why the oscillating effect appears in the position distributions. Of course, the results are completely the same with the results achieved by adopting $\bar{n}(x,t)\propto |\varepsilon E_{cr}g(x)f(t)|^2$ according to Refs.~\cite{Kravtcov:2018,Aleksandrov:2021ylw}. This illustrates again that our perspective picture and its interpretation are reliable. Particularly note that it can not only offer the exact location (position) of the created particle, but also explain the characteristics of the position distribution.
\begin{figure}[ht!]\centering
\includegraphics[width=0.5\textwidth]{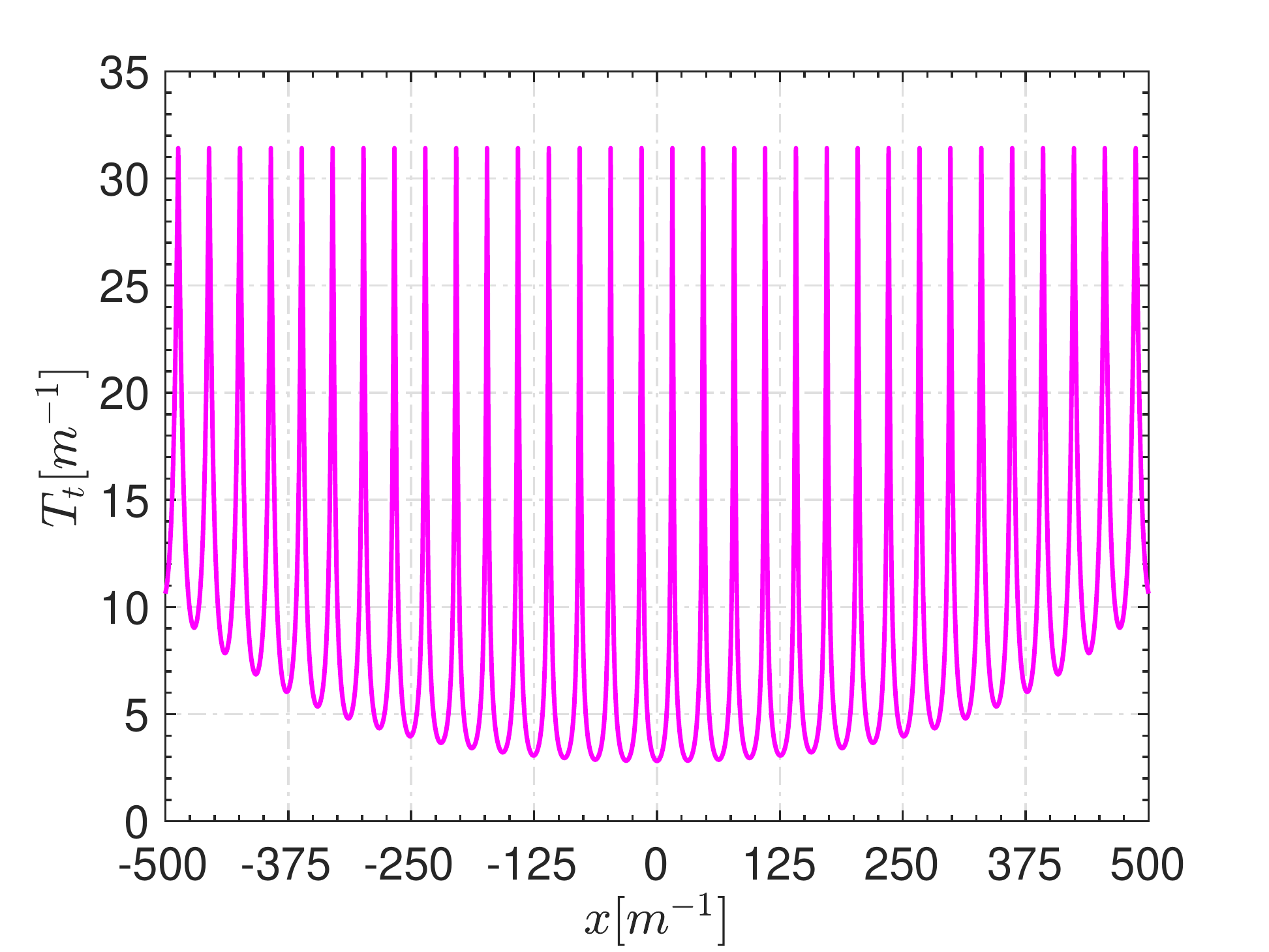}
\caption{Tunneling time for $k=0.1~m$, $\lambda=300~m^{-1}$ and $\omega=0.1~m$.
\label{fig:5}}
\end{figure}
We can also interpret the oscillating effect in Fig.~\ref{fig:4}(c) by using tunneling time. An example of the tunneling time is shown in Fig.~\ref{fig:5}, we can observe that an obvious oscillating effect of the tunneling time and its minima and maxima corresponds to $x=M\pi/k$ and $x=(1/2+M)\pi/k$, respectively, with an interval of $d=2\Delta x=\pi/k$. Since the particle number is inversely proportional the tunneling time~\cite{Amat:2022uxq}, we can find the selfconsistent oscillating effect in the position distribution in the Fig.~\ref{fig:4}(c). This can be regard as the another physical interpretation of the particle transition probability for every center-of-mass coordinate $x$. It should be pointed out that our new perspective picture provide us tunneling distance and the corresponding the tunneling time. On the contrary, the position of the created pair can be determined theoretically by the tunneling time. For instance, the position $x$ in Fig.~\ref{fig:5} is corresponding to the position $x$ in the position distribution plotting of Fig.~\ref{fig:4}(c).

\section{Summary}\label{sec:6}

Effect of spacetime-dependent spatially oscillating fields on the electron-positron pair production is studied numerically and analytically while the work is smaller than twice the electron mass. We further propose a new perspective picture for spatially oscillating fields when $W<2mc^2$. Under large spatial scale, we explain the characteristics of the position and momentum distribution through tunneling time, tunneling distance and energy gap between the positive and negative energy bands in the Dirac vacuum. Our results show that the maximum reduced particle number is about five times by comparing to maximum number for non-spatial oscillation. We find that the pair production results could be obtained by using LDA and AA when spatial oscillating cycle number is larger than two. Finally, we show relationship between the position distribution and tunneling time by employing the WI approach, and explain the position distribution.

\section{Acknowledgments}\label{sec:8}

Some helpful discussions with A. Ilderton and C. Kohlfuerst is acknowledged. This work was supported by the National Natural Science Foundation of China (NSFC) under Grant No. 11935008 and No. 12265024. The computation was carried out at the HSCC of the Beijing Normal University.

\end{document}